\documentclass[a4paper,12pt,reqno]{amsart}
\usepackage[latin1]{inputenc}
\usepackage[T1]{fontenc}
\usepackage{amsmath}
\usepackage{amsfonts}
\usepackage{amssymb}
\usepackage{amsthm}
\usepackage{color}
\usepackage{amsthm}
\usepackage{mathbbol}
\usepackage[citecolor=magenta,colorlinks=true]{hyperref}
\usepackage{graphicx}
\usepackage{mathrsfs, stmaryrd, mathtools}
\usepackage{cleveref}
\usepackage[style=ieee,citestyle=numeric-comp, backend=biber]{biblatex}
\DeclarePairedDelimiterX\intff[2]{[}{]}{#1,#2}
\DeclarePairedDelimiterX\intfo[2]{[}{)}{#1,#2}
\DeclarePairedDelimiterX\intof[2]{(}{]}{#1,#2}
\DeclarePairedDelimiterX\intoo[2]{(}{)}{#1,#2}

\DeclarePairedDelimiterX{\setof}[2]{\lbrace}{\rbrace}{#1\,{:}\,#2}
\DeclarePairedDelimiterX{\bracksof}[2]{[}{]}{#1\,\delimsize\vert\,#2}
\DeclarePairedDelimiterX{\parsof}[2]{(}{)}{#1\,\delimsize\vert\,#2}
\DeclarePairedDelimiterXPP\lnorm[2]{}\lVert\rVert{_{#1}}{#2}

\author[]{Debankur Das, Pappu Acharya, Kabir Ramola}

\thanks{\textsc{Centre for Interdisciplinary Sciences, Tata Institute of Fundamental Research, Hyderabad,  500046, Telengana, India}}

\thanks{\emph{E-mail:} \texttt{debankurd@tifrh.res.in}, \texttt{pappuacharya@tifrh.res.in}, \texttt{kramola@tifrh.res.in}}

\title[]{Displacement correlations in\\ disordered athermal networks}
\date{\today}

\begin{document}
\maketitle
\begin{abstract}
We derive exact results for correlations in the displacement fields $\{ \delta \vec{r} \} \equiv \{ \delta r_{\mu = x,y} \}$ in near-crystalline athermal systems in two dimensions. We analyze the displacement correlations produced by different types of microscopic disorder, and show that disorder at the microscopic scale gives rise to long-range correlations with a dependence on the system size $L$ given by $\langle \delta r_{\mu} \delta r_{\nu} \rangle \sim c_{\mu \nu}(r/L,\theta)$. In addition, we show that polydispersity in the constituent particle sizes and random bond disorder give rise to a logarithmic system size scaling, with $c_{\mu \nu}(\rho,\theta) \sim  \text{const}_{\mu\nu} - \text{a}_{\mu\nu}(\theta)\log \rho + \text{b}_{\mu\nu}(\theta) \rho^{2} $ for $\rho~(=r/L) \to 0$. This scaling is different for the case of displacement correlations produced by random external forces at each vertex of the network, given by $c^{f}_{\mu \nu}(\rho,\theta) \sim \text{const}^{f}_{\mu \nu} -( \text{a}^{f}_{\mu \nu}(\theta) + \text{b}^{f}_{\mu \nu}(\theta) \log \rho ) \rho^2 $. 
Additionally, we find that correlations produced by polydispersity and the correlations produced by disorder in bond stiffness differ in their symmetry properties.
Finally, we also predict the displacement correlations for a model of polydispersed soft disks subject to external pinning forces, that involve two different types of microscopic disorder. We verify our theoretical predictions using numerical simulations of polydispersed soft disks with random spring contacts in two dimensions.
\end{abstract}





\section{Introduction}\label{sec1}

Disordered networks~\cite{broedersz2014modeling} arise in several natural contexts and networks with soft potentials are paradigmatic systems in the study of elasticity~\cite{ostoja2002lattice}. In such systems, the microscopic architecture of the network is important in determining the mechanical response. However, theoretical frameworks are limited to continuum mechanical descriptions~\cite{vernerey2018statistical} or mean-field approximations~\cite{creton2016fracture}, which do not capture the role of structural disorder at the grain level. 
Recent studies have revealed that in the presence of thermal fluctuations~\cite{das2020wrinkles} the elastic properties of such networks is dependent on the rigidity of the constituents as well as temperature~\cite{dennison2013fluctuation,skrzeszewska2010fracture}.
However, as the size of the constituents is increased, the contributions arising from thermal fluctuations become negligible, and such systems are governed purely by their minimum energy configuration, and consequently by the constraints of mechanical equilibrium at the microscopic scale \cite{das2021long,acharya2020athermal,acharya2021disorder}. Such systems are therefore termed athermal, and the absence of thermal fluctuations leads to the system becoming arrested in the state of mechanical equilibrium, i.e. each particle is in force balance. 
Such situations naturally arise in jammed particle packings \cite{o2003jamming}, granular \cite{behringer2018physics,goldenberg2002force} and glassy systems \cite{ikeda}, active matter \cite{henkes}, as well as biological tissues \cite{dapeng1,broedersz2011criticality,boromand2018jamming}.
The complex random spatial geometry associated with energy minimized structures in such systems has made the development of coarse-grained theories challenging ~\cite{edwards1999statistical,edwards2001missing}.
Although, some characteristics can be predicted analytically ~\cite{degiuli2018field,lemaitre2014structural,maloney2006correlations,jishnu}, many of the properties such as the stability and response of such systems arising from the nature of the underlying disorder still remain elusive.


Near-crystalline systems provide a simpler arena to develop microscopic theories, as the perfect order associated with an ideal crystal makes it possible to calculate many properties analytically. Indeed, by introducing disorder into a crystalline system, it is also possible to move through the entire order-disorder spectrum \cite{tong2015crystals}. Large disorder can eventually lead to a completely amorphous regime. Disorder can be introduced into a crystalline system in many ways \cite{nie2017role, didonna2005nonaffine,ostoja2002lattice}. Some of the many examples include heterogeneity in the  properties of the network. Examples include introducing polydispersity in particle sizes~\cite{tong2015crystals,o2002random,tsekenis2021jamming,charbonneau2019glassy}, adding a random pinning force to each particle~\cite{schwarz2013physics,das2021long}, introducing topological disorder by randomly creating vacancies on some sites which leads to a departure from the lattice periodicity \cite{Goodrich2014}, randomly removing some bonds \cite{nie2017role}, and even spatial variation in the network geometry introduced through variable lengths of the bonds~\cite{thorpe1989rigidity}. Several studies have confirmed that uncorrelated microscopic disorder leads to highly correlated displacement fields \cite{das2021long, didonna2005nonaffine, ro2021disorder,degiuli2018field}. 
The correlations in such displacement fields produced by microscopic disorder can lead to an understanding of the local variation in the elastic properties and the deviation from crystalline behaviour. Therefore, studying the displacement correlations and their symmetry properties is a natural way to understand the emergence of elastic behaviour in the presence of disorder. 
The introduction of a small disorder in the system enables us to  analytically predict various properties exactly from the first principle calculations \cite{ acharya2021disorder,taraskin2001origin, elliott1974theory}. Networks with random bond stiffness in the low disorder regime are often used to model the origin of the boson peak in the low frequency vibrational density of states of amorphous systems~\cite{taraskin2001origin, schirmacher1998harmonic, gratale2013phonons,kantelhardt2002vibrational}. Elastic networks with random bond stiffness are also encountered in understanding the rheological properties of soft solids~\cite{de2020viscoelastic}. 


In this paper, we derive the correlations in the displacement fields produced by multiple types of microscopic disorder, polydispersity in particle sizes, randomness in the bond stiffness as well as pinned external forces at each vertex. Our formalism makes no appeal to a coarse-graining procedure~\cite{didonna2005nonaffine}, instead directly utilizes the disorder at the grain level to derive the exact form of the displacement fields. This leads to a microscopic derivation of displacement fields on a disordered background, that are not amenable to a continuum elasticity treatment. The framework developed in this paper can therefore be used as an input for other studies that require the details of the geometric disorder in such networks.
Our analysis shows that the long-range correlations arising from disorder in bond stiffness differ in symmetry properties from correlations produced by polydispersity. We provide analytic predictions for these correlations for multiple types of microscopic disorder and verify them with numerical simulations. We also include external disorder by incorporating random pinning or active forces at each vertex of the network. In addition, we show that the correlations arising due to polydispersity as well as the presence of external forces at each vertex and can be determined as a superposition of the individual correlations.

The outline of this paper is as follows. In Section ~\ref{sec_model}, we provide a detailed description of the disordered network model with microscopic disorder arising from random bond stiffness between particles, polydispersity in particle sizes as well as random pinning forces at each vertex.
In Section ~\ref{sec_method} we review the perturbation expansion method developed in earlier work \cite{acharya2021disorder,acharya2020athermal,das2021long} and provide the general expression for the correlations in the displacement fields for different types of quenched disorder. In Section~\ref{sec_bond_disorder}, we derive the theoretical form of displacement correlations in the presence of bond disorder. 
We also analytically predict the system size dependence of the correlation functions in the continuum limit. In Section~\ref{sec_correlations_poly}, we compute the displacement correlations produced by the presence of polydispersity in particle sizes, which display different symmetry properties compared to those arising from bond disorder. We derive the continuum limit of these correlation functions and compare our results with numerical simulations. Our theoretical framework in Section~\ref{sec_bond_disorder} and Section~\ref{sec_correlations_poly} shows that both these correlations display long-range behaviour. In Section~\ref{sec_correlations_random_forces}, we revisit the theoretical framework used to derive displacement correlations when external pinning forces are applied to each site \cite{das2021long}. The long-range nature of these displacement correlations is further investigated and found to be fundamentally different from the displacement correlations arising from bond disorder and particle size polydispersity. Further, in Section~\ref{sec_correlations_both}, we show that the displacement correlations in a polydispersed athermal membrane with external pinning forces at each site can be obtained as the linear combination of the correlations arising due to each disorder individually. Finally, we conclude by discussing the relevance of our results and providing avenues of future research in Section~\ref{sec_conclusions}.  

\section{Disordered Network Model}\label{sec_model}

\begin{figure}[t!]
\centering
\hspace{-2cm}
\includegraphics[width=1.0\linewidth]{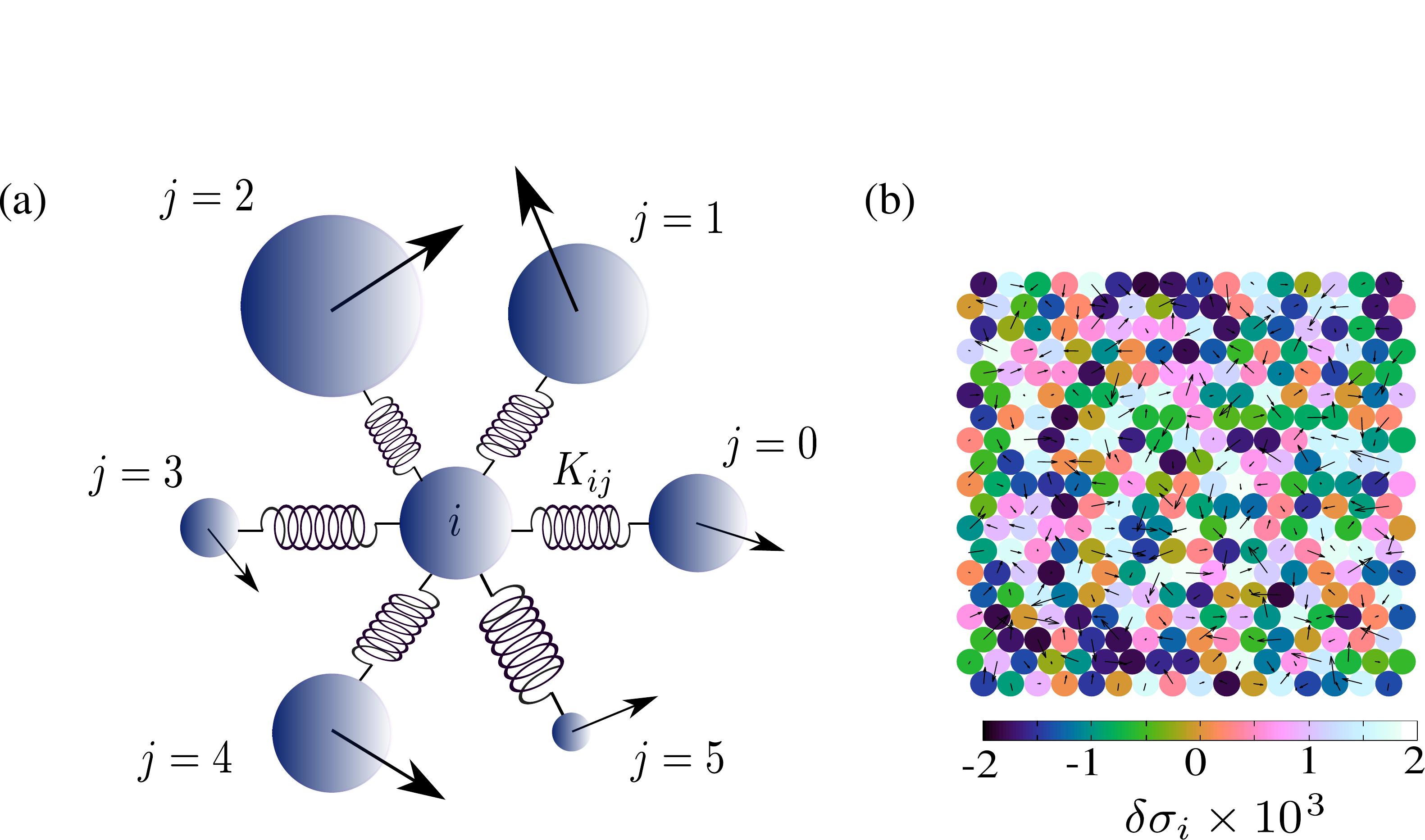}
\caption{
\textbf{(a)} Schematic representation of the network model, a triangular arrangement of polydispersed soft disks with random bond stiffness between the particles. Three possible types of disorder are introduced in the system (i) polydispersity in particle sizes represented by differences in particle radii $\sigma_i$, (ii) disorder in the bond stiffness of neighbouring particles $K_{ij}$ and (iii) disorder due to the presence of random forces at each point in the lattice network (represented by arrows). \textbf{(b)}  A disordered energy-minimized configuration of polydispersed soft disks (the colors represent the incremental sizes of the particles) with pinning forces (represented with arrows) applied on each particle.}\label{fig_main_schematic}
\end{figure}

We analyze a system of particles arranged in an $L \times L$ triangular lattice grid (see Fig.~\ref{fig_main_schematic}(b)) with periodic boundary conditions. The initial crystalline position of each particle is denoted by $\{\vec{r}_{i}^{(0)}\} \equiv \{x_{i}^{(0)},y_{i}^{(0)}\}$.
The particles interact with their nearest neighbours through soft potentials, in which we incorporate microscopic disorder through randomness in the sizes, as well as the bond stiffness.
The total energy of the system is given by
\begin{eqnarray}
\nonumber
\mathcal{H} &=&  \sum_{i=1}^{L^2} \sum_{\langle i j \rangle} \frac{K_{ij}}{\alpha}\Big (1 - \frac{\mid \vec{r}_{ij} \mid}{\sigma_{ij}} \Big)^\alpha,~~  \mid r_{ij}\mid ~~\le \sigma_{ij}, \\
&=& 0, \hspace{44mm} \mid r_{ij}\mid ~~> \sigma_{ij},
\label{eq_hamiltonian}
\end{eqnarray}
where $\vec{r}_{ij} =  \vec{r}_{j} - \vec{r}_{i}$ represents the distance between the $i^{\text{th}}$ and $j^{\text{th}}$ particle and $\sigma_{ij} = (\sigma_{i} + \sigma_{j})$, and $\sigma_i$ and $\sigma_j$ are the radii of the particles. $K_{ij}$ denotes the spring constant between the particles $i$ and $j$. The brackets $\langle \rangle$ denote nearest-neighbours on the network. When all the spring constants are equal, the interaction in Eq.~(\ref{eq_hamiltonian}) reduces to the paradigmatic soft sphere Hamiltonian \cite{o2003jamming,tong2015crystals,Goodrich2014,tsekenis2021jamming,o2002random} used to model frictionless materials such as foams and colloids. 
The soft sphere model fails to capture the interactions of non-spherical shapes, as well as their elastic heterogeneity. The variation in the spring constants $K_{ij}$ can therefore provide a way to add randomness in the shapes and elastic moduli of the particles. When all the radii are set equal, the model reduces to another paradigmatic form: a crystalline network with random bond stiffness~\cite{taraskin2001origin, schirmacher1998harmonic, gratale2013phonons,kantelhardt2002vibrational,de2020viscoelastic}.
In this work, we set $\alpha = 2$ (harmonic), however our techniques can be trivially generalized to any value of $\alpha$. The initial crystalline arrangement consists of particles with equal radii $\sigma_0$ and the interparticle spring constant is $K_{ij} = 1$. The lattice constant is $R_0 = 2 \sigma_{0}(1 - \varepsilon)$. Here $\varepsilon$ quantifies the overcompression of the initial crystalline state given by $\varepsilon = 1 - \sqrt{\phi_c/\phi}$, where $\phi$ is the packing fraction of the system and $\phi_c = \pi/\sqrt{12}$ is the packing fraction of the hexagonal close packed crystal (i.e. $\varepsilon = 0$) with no overlaps.

In this paper, we study three different limits of the generalized model presented above, that arise in various physical situations.  

(i) {\bf Triangular lattice with random bond stiffness}, where, disorder in the lattice network is introduced by applying a small randomness in the interparticle bond stiffness $K_{ij} = K_0 +\delta K_{ij}$.  This is a useful model for various biological systems where different constituent proteins exhibit different types of hydrogen bonding~\cite{hildebrand2008hydrogen}. Such bond disorder is also encountered in the study of alloys and intercalants~\cite{mousseau1993structural,thorpe1989rigidity}.

(ii) {\bf Polydispersed soft particle crystals}, where we introduce disorder (polydispersity) in the radii of individual particles $\sigma_i = \sigma_0 + \delta \sigma_i$. This is a fundamental type of disorder that arises naturally in jammed athermal systems 
\cite{o2002random,tong2015crystals,acharya2020athermal}. The variation in particle sizes plays a significant role in determining the bulk properties as well as the dynamics of the crystalline and amorphous phases.  

(iii) {\bf Triangular network with random pinning forces}. In addition to inherent randomness in the underlying structure of the system, we consider a system with externally imposed forces, which provide an external source of randomness. The presence of such disorder is studied extensively in our earlier work ~\cite{das2021long}, where we computed the displacement fields produced by pinning forces and also identified the long-range nature of the displacement correlations. Here, we extend our earlier model by imposing polydispersity in the particle sizes along with the randomly pinned external forces.

(iv) {\bf Polydispersed crystals with random pinning}. Finally we study the effect of multiple types of disorder in the system and show that they can be well-described by the superposition of the individual correlations. We illustrate this with an example of a polydispersed crystal with random pinning forces where the interplay between the different symmetries of the underlying correlation functions can produce non-trivial lengthscales in the system.







\subsection*{Numerical Simulations}
In order to verify the predictions from our theory, 
we simulate an athermal triangular network in the presence of various types of disorder. In our simulations we consider jammed states, with each particle in force balance, i.e. configurations at energy minima. The energy of the system is minimized using the FIRE (Fast Inertial Relaxation Engine) algorithm \cite{fire}, which can easily incorporate both inherent microscopic disorder and external forces. In this study, we have chosen a packing fraction of $\phi = 0.92$, which fixes the interparticle distance in the reference crystal to be $R_0 = 0.99285$. The value of reference bond stiffness is kept fixed at $K_0 = 1$ and the initial radii of the particles is set to $\sigma_0 = \frac{1}{2}$. The numerical results provided in this paper are averaged over 1000 random configurations for three different system sizes $L = 32,64,128$.
 

\section{Linearized Perturbation Expansion}
\label{sec_method}

In earlier  work~\cite{das2021long,acharya2021disorder,acharya2020athermal}, we have developed a perturbative framework to calculate the displacement fields up to arbitrary accuracy in an athermal network in the presence of disorder. This framework is applicable for general disorder of small magnitude and for systems of particles interacting via pairwise central potentials given by
\begin{eqnarray}
u(\vec{r}_{ij},\{\zeta_{i}\}) = \mathcal{F}\left(| \vec{r}_{ij}|,\{\zeta_{i}\}\right).
\label{general_energy_law}
\end{eqnarray}
Here $\vec{r}_{ij} = \vec{r}_{j} - \vec{r}_{i}$ is the vector distance between the particles $i$ and $j$ located at positions $\vec{r}_i$ and $\vec{r}_j$ respectively. In this paper, we consider the pairwise central potential to be short ranged. The variables $\{ \zeta_i \}$ represent quenched scalar variables at every site which can also be assigned to each bond, and can be tuned to create perturbations about the crystalline state. As a response to the microscopic disorder, the positions of the particles change from their crystalline positions as
\begin{eqnarray}
\nonumber
x_{i} &=& x_{i}^{(0)} + \delta x_{i},\\
y_{i} &=& y_{i}^{(0)} + \delta y_{i}.
\end{eqnarray}
These displacement fields in real space $\delta x_i \equiv \delta x(\vec{r})$ and $\delta y_i \equiv \delta y(\vec{r})$ are assigned at each site $i \equiv \vec{r}_i \equiv \vec{r}$ of the {\it reference} crystal structure.
The perturbed energy minimized configuration has contact forces $\{ \vec{f}_{ij} \}$ between particles $i$ and $j$ such that mechanical force balance is maintained for each particle i.e. 
\begin{equation}
\sum_{j} f^{x}_{ij} = 0, ~~~~~\sum_{j} f^{y}_{ij} = 0, ~~~~\forall ~i.
\label{eq_force_balance_full}
\end{equation}
Here $f^{x(y)}_{ij}$ are the $x(y)$ components of the forces between particles $i$ and $j$, and the sum is over all particles $j$ in contact with particle $i$. 
The displacement fields produced by disorder lead to modified interparticle forces, which can be computed as a perturbation expansion of the Hamiltonian in Eq.~\eqref{eq_hamiltonian}. Imposing the force balance conditions in Eq.~\eqref{eq_force_balance_full} leads to closure equations for the displacement fields, which can then be computed within a heirarchical perturbation expansion \cite{acharya2021disorder}.
Although this perturbative method can be used to calculate the displacement fields up to any order, we compute the displacement correlations using only the linear order solutions which provides the leading order behaviour. At linear order, the force balance conditions in Eq.~\eqref{eq_force_balance_full} can be expressed as
\begin{eqnarray}
\nonumber
\sum_j \left(C_{ij}^{xx}\delta x^{}_{ij} + C_{ij}^{xy}\delta y^{}_{ij} + \Omega_{x}(\delta \zeta_{ij}) \right) &=& 0, ~\forall ~i, \\
\sum_j \left(C_{ij}^{yx}\delta x^{}_{ij} + C_{ij}^{yy}\delta y^{}_{ij} + \Omega_{y}(\delta \zeta_{ij})\right) &=& 0, ~\forall ~i.
\label{all_order_force_balance}
\end{eqnarray}
where the $\Omega_{x}$ and $\Omega_{y}$ terms are composed of microscopic disorder and depend on the magnitude of the disorder at sites $i$ and $j$ and also on the symmetries of the underlying lattice. 
The linear coefficients $C_{ij}^{\mu \nu}$ are equivalent to the elements of the Hessian matrix \cite{acharya2021disorder} of the system and can be expressed purely in terms of the reference crystalline state.
Crucially, the system of equations in Eq.~\eqref{all_order_force_balance} can be inverted in Fourier space~\cite{acharya2021disorder}. The displacement fields at first order can be expressed in Fourier space as ~\cite{das2021long}
\begin{eqnarray}
\nonumber
\delta \tilde{x}(\vec{k})&=& \left(\tilde{G}^{xx}(\vec{k}) \tilde{S}^{x}(\vec{k}) + \tilde{G}^{xy}(\vec{k}) \tilde{S}^{y}(\vec{k}) \right),\\
\delta \tilde{y}(\vec{k}) &=& \left(\tilde{G}^{yx}(\vec{k}) \tilde{S}^{x}(\vec{k}) + \tilde{G}^{yy}(\vec{k}) \tilde{S}^{y}(\vec{k}) \right).
\label{eq_disp_fourier}
\end{eqnarray}
Here $\delta \tilde{x} (\vec{k}) = \sum_{\vec{r}} \exp(i \vec{k}. \vec{r}) \delta x (\vec{r})$ and $\delta \tilde{y} (\vec{k}) = \sum_{\vec{r}} \exp(i \vec{k}. \vec{r}) \delta y (\vec{r})$ represent the Fourier transform of the real space displacement fields $\delta x(\vec{r})$ and $\delta y(\vec{r})$ respectively. $\tilde{S}^{x}(\vec{k})$ and $\tilde{S}^{y}(\vec{k})$ are the linear order source terms in Fourier space, composed of the microscopic disorder $\Omega_{x}(\delta \zeta_{ij})$ and $\Omega_{y}(\delta \zeta_{ij})$. The Green's functions $\tilde{G}^{\mu\nu}$ in Fourier space are given by \cite{das2021long}
\begin{eqnarray}
\begin{aligned}
\tilde{G}^{xx}(\vec{k}) = \Gamma_1 \frac{ (1-\epsilon)}{\Gamma_1\Gamma_2 -\Gamma^2_3} ,\\
\tilde{G}^{xy}(\vec{k}) = \Gamma_3 \frac{ (1-\epsilon)}{\Gamma_1\Gamma_2 -\Gamma^2_3}, \\
\tilde{G}^{yx}(\vec{k}) = \Gamma_3 \frac{ (1-\epsilon)}{\Gamma_1 \Gamma_2 -\Gamma^2_3}, \\
\tilde{G}^{yy}(\vec{k}) = \Gamma_2 \frac{ (1-\epsilon)}{\Gamma_1 \Gamma_2 -\Gamma^2_3}, 
\label{eq_exact_green}
\end{aligned}
\end{eqnarray}
where the functions $\Gamma_1$, $\Gamma_2$, $\Gamma_3$ have the following forms
\begin{eqnarray}
\nonumber
\Gamma_1 &=& (3-4 \epsilon ) \cos (k_x) \cos (k_y)-2 \epsilon  \cos (2 k_x)+6 \epsilon -3,\\
\nonumber
\Gamma_2 &=& (1-4 \epsilon ) \cos (k_x) \cos (k_y)+2 ( 1-\epsilon) \cos (2 k_x)+6 \epsilon -3,\\
\Gamma_3 &=&\sqrt{3}\sin(k_x)\sin(k_y).
\end{eqnarray}
The displacement fields in real space can then be obtained by taking an inverse Fourier transform. We have
\begin{small}
 \begin{eqnarray}
\nonumber
 \delta x^{}(\vec{r}) &=& \sum_{\vec{r}^\prime} G^{xx}(\vec{r}- \vec{r}^\prime)S^{x}(\vec{r}^\prime) + \sum_{\vec{r}^\prime} G^{xy}(\vec{r}- \vec{r}^\prime)S^{y}(\vec{r}^\prime),\\
 \delta y^{}(\vec{r}) &=& \sum_{\vec{r}^\prime} G^{yx}(\vec{r}- \vec{r}^\prime)S^{x}(\vec{r}^\prime) + \sum_{\vec{r}^\prime} G^{yy}(\vec{r}- \vec{r}^\prime)S^{y}(\vec{r}^\prime).
\label{eq_norder_disp_gen}
\end{eqnarray}
\end{small}
Here, $G^{\mu\nu}(\vec{r} - \vec{r}^\prime)$ are the real space Green's functions and can be computed as an inverse Fourier transform of the expressions in Eq.~(\ref{eq_exact_green}) as $G^{\mu \nu}(\vec{r}) = \frac{1}{V}\sum_{\vec{k}}\tilde{G}^{\mu \nu}(\vec{k})e^{-i\vec{k}.\vec{r}}$.
Here $V = 2L^2$ is the volume of the system, following the convention of earlier work~\cite{das2021long,acharya2021disorder,horiguchi}.
We note that the displacement fields arising from different microscopic disorders in Eq.~\eqref{eq_norder_disp_gen} involve the same Green's functions, and only modify the source terms.

The fundamental quantity of interest in this paper is the correlations in the displacement fields in real space. Using Eq.~(\ref{eq_norder_disp_gen}), the displacement correlations can be expressed as 
\begin{equation}
\mathcal{C}_{\mu \nu}(\vec{r} - \vec{r}^{\prime}) = \langle \delta r_{\mu}(\vec{r}) \delta r_{\nu} (\vec{r}^{\prime}) \rangle.
\end{equation}
The real space displacement correlations can be written in terms of the displacements in Fourier space as
\begin{eqnarray}
\hspace{-0.8cm}
\langle \delta r_{\mu}(\vec{r}) \delta r_{\nu}(\vec{r}^\prime) \rangle = \Big\langle \frac{1}{V^2} \sum_{\vec{k},\vec{k}^\prime} \delta r_{\mu} (\vec{k}) \delta r_{\nu}(\vec{k}^\prime) \exp(-i \vec{k} \cdot \vec{r}) \exp(-i \vec{k}^\prime \cdot \vec{r}^\prime) \Big \rangle,
\label{eq_correlation1}
\end{eqnarray}
where $\{ \delta \vec{r} \} \equiv \{ \delta r_{x}, \delta r_{y} \} \equiv \{ \delta x, \delta y \}$.
Using the displacement fields at linear order and inserting Eq.~(\ref{eq_disp_fourier}) in Eq.~(\ref{eq_correlation1}), we obtain the displacement correlations in the real space in terms of the sources. We have
\begin{eqnarray}
\hspace{-1cm}
\langle \delta r_{\mu}(\vec{k}) \delta r_{\nu}(\vec{k}^\prime) \rangle &=& \sum_{\alpha \equiv x,y} \sum_{\beta \equiv x,y}  \tilde{G}^{\mu \alpha}(\vec{k}) \tilde{G}^{\nu \beta} (\vec{k}^\prime) \langle \tilde{S}^{\alpha}(\vec{k}) \tilde{S}^{\beta}(\vec{k}^\prime) \rangle.
\label{eq_correlations_exp_K}
\end{eqnarray}
The translation invariance of the correlation function requires that the expressions in Eq.~(\ref{eq_correlations_exp_K}) are non-zero only when $\vec{k} = -\vec{k}^\prime$.  
The quantity $\langle \tilde{S}^{\alpha}(\vec{k}) \tilde{S}^{\beta}(\vec{k}^\prime) \rangle$ is the correlation of the Fourier transform of the source terms which we use to compute the correlations in the displacement fields through Eq.~\eqref{eq_correlations_exp_K}. In the subsequent sections, we study displacement correlations in physical systems in the presence of various types of disorder, which involve different source terms. Further, using the limit $|k| \to 0$ and converting the summation in Eq.~(\ref{eq_correlation1}) to an integral, we can also derive the continuum limit of these correlations and predict the scaling forms, as  described in this paper.

\section{Triangular Lattice with Random Bond Stiffness}
\label{sec_bond_disorder}
We begin by using the general formulation developed in Section \ref{sec_method} to analyze the displacement correlations in a triangular arrangement of particles with random bond stiffness. The particles sizes are held fixed as $\sigma_{i} =  \sigma_0 = 1/2$. 
The randomness in the interparticle interaction is incorporated by introducing disorder in the individual bond stiffness with magnitudes $K_{ij}$, which are chosen to be independent random variables at each bond. The bond stiffness between particles $i$ and its $j^{th}$ neighbour assigned to each site $(x_i,y_i)$ is given by
\begin{eqnarray}
K_{ij} (x_i,y_i) = K_0(1 + \eta_{K} \xi_{i}) =  K_0 + \delta K_{ij},
\end{eqnarray}
where $\xi_i$ are independent random variables at each bond, which we choose to be drawn from a uniform distribution between $[-1/2,1/2]$. For convenience, we set $K_0 = 1$ in our computations. Here $\eta_{K}$ represents the strength of the bond disorder. The correlations in the bond stiffness $\delta K_{ij} (x_i,y_i)$ are then
\begin{small}
\begin{eqnarray}
\nonumber
&&\langle \delta K_{ij}(x_i,y_i) \delta K_{i^{\prime} j^{\prime}}(x_i^\prime, y_i^\prime) \rangle\\
\nonumber
=&&\frac{\eta_{K}^2}{12} K_0 \Big [ \delta \left(x_{i}^{\prime(0)} - x_{i}^{(0)} \right) \delta \left(y_{i}^{\prime(0)} - y_{i}^{(0)} \right) \delta_{j^\prime j} +\delta \left (x_{i}^{\prime(0)} - x_{i}^{(0)} - R_{0} \cos \left(\frac{2\pi j}{3} \right) \right)\\
&&\hspace{2.5cm} \times \delta \left(y_{i}^{\prime(0)} - y_{i}^{(0)} - R_{0} \sin\left(\frac{2\pi j}{3}\right) \right) \delta_{j^\prime, (j+3)\bmod 6} \Big],
\label{eq_correlation_K}
\end{eqnarray}
\end{small}
where $\vec{r}_{i}^{(0)} \equiv {(x_i^{(0)},y_i^{(0)})}$ represents the coordinate of the $i^{th}$ particle in the initial reference lattice.
The somewhat complicated form of the above correlation occurs because our convention counts each bond twice (once for each particle). Each bond is labeled twice: as the $j$th bond of the particle at $\vec{r}_i$ as well as the $j'$th bond of the neighbouring particle at $\vec{r}_i'$. This leads to the second term in the expressions in Eq.~(\ref{eq_correlation_K}). Using Eq.~(\ref{all_order_force_balance}), the force balance condition at each vertex of the network can be written as
\begin{eqnarray}
\nonumber
\sum_j \Big(C_{ij}^{xx}\delta x^{}_{ij} + C_{ij}^{xy}\delta y^{}_{ij} + C_{ij}^{xK} \delta K_{ij} (\vec{r}_i) \Big) &=& 0, ~\forall ~i, \\
\sum_j \left(C_{ij}^{yx}\delta x^{}_{ij} + C_{ij}^{yy}\delta y^{}_{ij} + C_{ij}^{yK} \delta K_{ij} (\vec{r}_i) \right) &=& 0, ~\forall ~i.
\label{all_order_force_balance_K}
\end{eqnarray}
where $C_{ij}^{xK} = -2 \sigma_0 \epsilon\cos(\frac{\pi j}{3})$ and $C_{ij}^{yK} = -2\sigma_0 \epsilon \sin(\frac{\pi j}{3})$. 
Above, the disorder in the bond stiffness act as source terms that generate the displacement fields through the microscopic force balance conditions.

\begin{figure}[t!]
\centering
\includegraphics[width=0.9\linewidth]{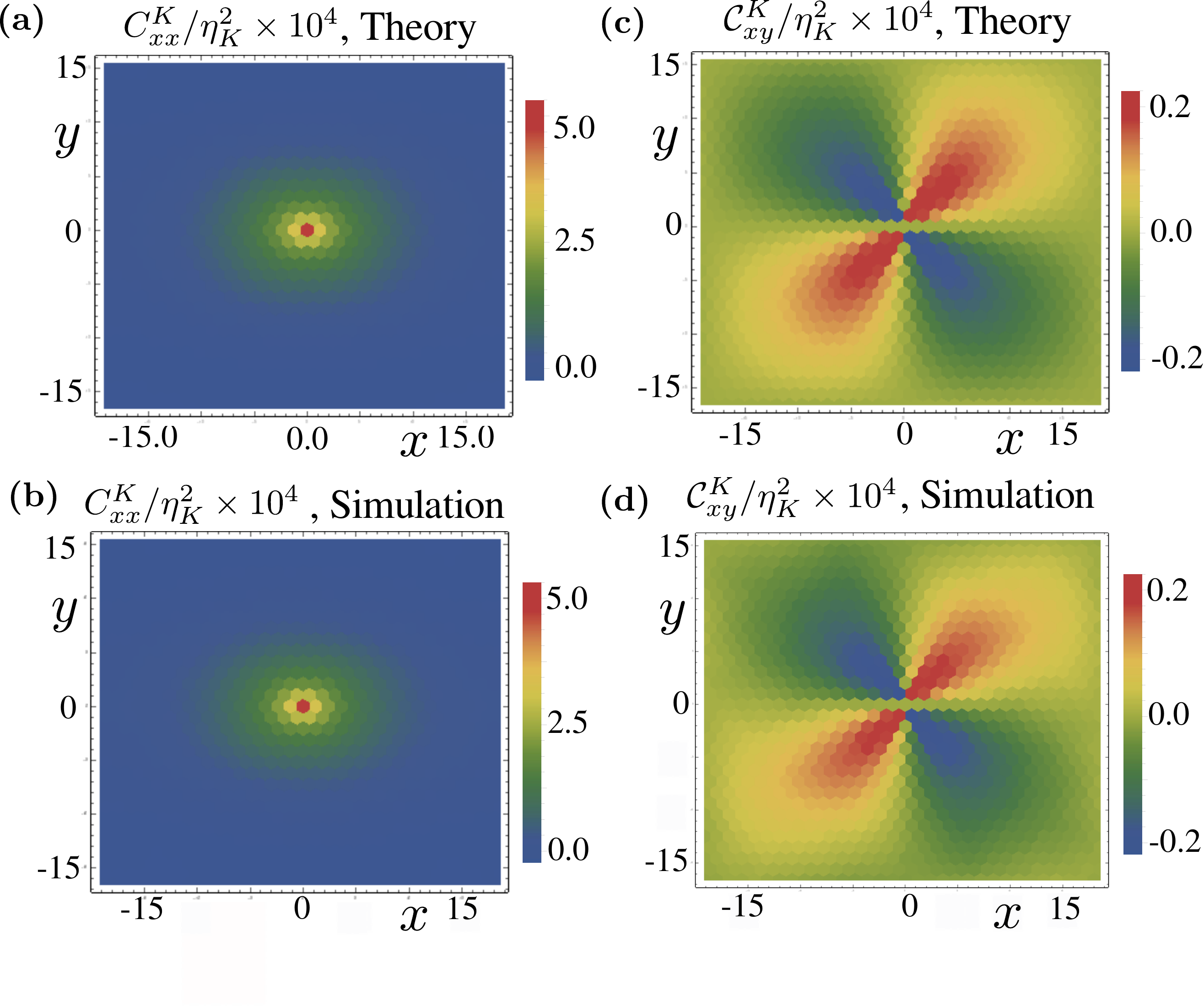}
\caption{Correlations in the displacement fields produced by randomness in the stiffness constant of the interparticle bonds. \textbf{(a)} $\mathcal{C}^{K}_{xx}(\vec{r}-\vec{r}' ) = \langle \delta x(\vec{r}) \delta x( \vec{r}' ) \rangle$ obtained from the expressions in Eq.~\eqref{eq_correlation_K_final}. \textbf{(b)} $\mathcal{C}^{K}_{xx}(\vec{r}-\vec{r}' ) = \langle \delta x(\vec{r}) \delta x( \vec{r}' ) \rangle$ obtained from the simulations. \textbf{(c)} $\mathcal{C}^{K}_{xy}(\vec{r}-\vec{r}' ) = \langle \delta x(\vec{r}) \delta y( \vec{r}' ) \rangle$ obtained from the expressions in Eq.~\eqref{eq_correlation_K_final}. \textbf{(d)} $\mathcal{C}^{K}_{xy}(\vec{r}-\vec{r}' ) = \langle \delta x(\vec{r}) \delta y( \vec{r}' )\rangle$ obtained from simulations. The magnitude of the disorder used is $\eta_K = 0.001$.
}
\label{fig_corr_K}
\end{figure}

The source terms at each site associated with the bond disorder 
can then be expressed as
\begin{eqnarray}
\nonumber
{S}^{x K}_i(\vec{r}_{i}) &=& \sum_j C_{ij}^{xK} \delta K_{ij} (\vec{r}_{i}), \\
{S}^{y K}_i(\vec{r}_{i}) &=& \sum_j C_{ij}^{yK} \delta K_{ij} (\vec{r}_{i}).
\end{eqnarray}
In order to calculate the displacement correlations using Eq.~(\ref{eq_correlations_exp_K}), we first compute the  correlations of the source terms in Fourier space 
\begin{small}
\begin{eqnarray}
\hspace{-0.7cm}
\langle \tilde{S}^{\alpha K}(\vec{k}) \tilde{S}^{\beta K}(\vec{k}^\prime) \rangle = \sum_{j,j^\prime} C_{ij}^{\alpha K} C_{i^\prime j^\prime}^{\beta K}\langle \delta K_{ij}(\vec{r}_{i}) \delta K_{i^\prime j^\prime}(\vec{r_i^\prime}) \rangle \exp(i \vec{k} \cdot \vec{r} + i \vec{k}^\prime \cdot \vec{r}^\prime).
\end{eqnarray}
\end{small}
Next, using the form of the bond stiffness correlations in Eq.~(\ref{eq_correlation_K}) in Eq.~(\ref{eq_correlations_exp_K}), the displacement correlations can be expressed as
\begin{eqnarray}
\hspace{-0.5cm}
\tilde{\mathcal{C}}^{K}_{\mu \nu} (\vec{k}) 
&=& \frac{\eta_K^2}{12}\Big[\sum_{\alpha,\beta} \left(\tilde{G}^{\mu \alpha}(\vec{k}) \tilde{G}^{\nu \beta}(-\vec{k}) \right ) \left ( \mathcal{S}^{\alpha \beta}_{K1}(\vec{k}) + \mathcal{S}^{\alpha \beta}_{K2}(\vec{k}) \right) \Big],
\label{eq_correlation_K_finalk}
\end{eqnarray}
where the source terms $\mathcal{S}^{\alpha \beta}_{K1}(\vec{k})$ and $\mathcal{S}^{\alpha \beta}_{K2}(\vec{k})$ are translationally invariant and are given by
\begin{eqnarray}
\nonumber
\mathcal{S}^{\alpha \beta}_{K1}(\vec{k}) &=& \sum_{j=0}^5 C_{ij}^{\alpha K} C_{ij}^{\beta K},\\
\mathcal{S}^{\alpha \beta}_{K2}(\vec{k}) &=& \sum_{j=0}^5 C_{ij}^{\alpha K} C_{i(j + 3)\bmod 6}^{\beta K}  \exp (\vec{k} \cdot \vec{\mathbb{r}}_j).
\label{eq_correlation_K_mod}
\end{eqnarray}
Here, $\vec{\mathbb{r}}_j =  (R_{0}\cos(\pi j/3), R_{0}\sin(\pi j/3))$ are the fundamental lattice translation vectors.
The correlations in real space can then be extracted by performing an inverse Fourier transform of Eq.~(\ref{eq_correlation_K_finalk}). We have
\begin{eqnarray}
\nonumber
\mathcal{C}^{K}_{\mu \nu} (\vec{r} - \vec{r}^\prime) &=& \langle \delta r_{\mu}(\vec{r}) \delta r_{\nu}(\vec{r}') \rangle \\
 \nonumber
&=& \frac{\eta_K^2}{12} \sum_{\vec{k}} \sum_{\alpha,\beta} \left(\tilde{G}^{\mu \alpha}(\vec{k}) \tilde{G}^{\nu \beta}(-\vec{k}) \right ) \left ( \mathcal{S}^{\alpha \beta}_{K1}(\vec{k}) + \mathcal{S}^{\alpha \beta}_{K2}(\vec{k}) \right) \\
&&\hspace{3cm} \times \exp[-i \vec{k}\cdot (\vec{r} -\vec{r}^\prime)],
\label{eq_correlation_K_final}
\end{eqnarray}
In Fig.~\ref{fig_corr_K} ({\textbf a}) and Fig.~\ref{fig_corr_K} ({\textbf c}) we plot the theoretically computed correlations $\mathcal{C}_{xx}^K(\vec{r})$ and  $\mathcal{C}_{xx}^K(\vec{r})$ respectively. We also plot the numerically generated correlations obtained by averaging over 1000 realizations of the bond disorder (with $\eta_K = 0.001$) in Fig.~\ref{fig_corr_K} ({\textbf b}) and Fig.~\ref{fig_corr_K} ({\textbf d}). The numerically generated correlations match with our theoretical results exactly.

\subsection*{Continuum Limit}
Next, using Eq.~(\ref{eq_correlation_K_final}), we calculate the continuum form of the correlation in the limit $|k| \to 0$. 
The Green's function has a simple form in this limit $\Tilde{G}^{\mu \nu}(\vec{k}) = \frac{\tilde{g}^{\mu \nu}(\psi)}{k^2}$. In the limit $|k| \to 0$ the term $\mathcal{S}_{K2}^{\alpha \beta}(\vec{k}) \sim -\mathcal{S}_{K1}^{\alpha \beta}(\vec{k}) + s_{K2}^{\alpha \beta}(\psi){k^2}$ and hence the total source terms in Fourier space can be written as $\Big(\mathcal{S}_{K1}^{\alpha \beta}(\vec{k}) + \mathcal{S}_{K2}^{\alpha \beta}(\vec{k})\Big) \simeq s_{K2}^{\alpha \beta}(\psi){k^2}$. In the $L \to \infty$ limit, we can transform the sum in Eq.~\eqref{eq_correlation_K_final} into an integral and the displacement correlations can be expressed as 
\begin{eqnarray}
\nonumber
\mathcal{C}^{K}_{\mu \nu }(\vec{r}) &=& \frac{\eta_{K}^{2}}{12(2 \pi)^2} \int_{-\pi}^{\pi} \sum_{\alpha,\beta}\big[(\Tilde{g}^{\mu \alpha}(\psi)\Tilde{g}^{\nu \beta}(\psi) (s_{K2}^{\alpha \beta})(\psi) \big] d\psi\\
&&\hspace{3cm} \times \underbrace{{\int_{\lambda}^{\pi}\frac{\exp( -{i \vec{k}.\vec{r}})}{k} d{k}.}}_{\mathcal{I}^{K}(\lambda,r,\theta,\psi)}
\label{eq_correlation_K_intermediate}
\end{eqnarray}

\begin{figure}[t!]
\centering
\includegraphics[width=0.9\linewidth]{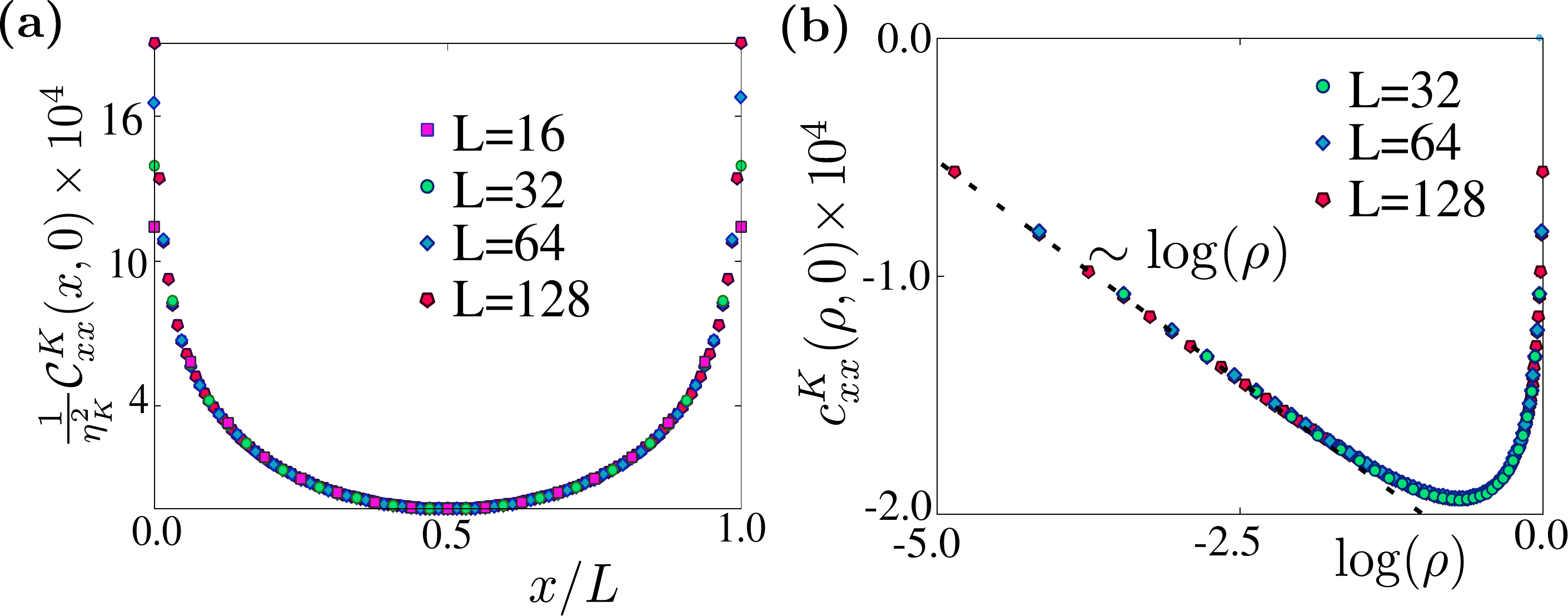}
\caption{\textbf{(a)} $\delta x$ correlations produced by bond disorder, measured along the $x$ direction $\mathcal{C}^{K}_{xx}(x,0) = \langle \delta x(\vec{r}) \delta x(\vec{r}+x\hat{x} ) \rangle$, for different system sizes. These correlations follow the scaling prediction in Eq.~(\ref{eq_corr_K}), with a scaling variable $\rho = r/L$. \textbf{(b)} These correlations display logarithmic behaviour as $\rho \to 0$ consistent with our predictions $\mathcal{C}^{K}_{\mu \nu}(\vec{r}) \sim \log(\rho)$ in Eq.~(\ref{eq_scaling_collapse_K}).
}
\label{fig_corr_K1}
\end{figure}
In performing the integral over the radial coordinate in Fourier space, we exclude the singular point $|\vec{k}| \to 0$. This is achieved by defining a cutoff $\lambda$ that represents a system-size dependent lower limit in radial coordinates in Fourier space as $\lambda = \frac{2\pi}{L}\Lambda$. Here $\Lambda$ represents an $\mathcal{O}(1)$ tuning parameter that accounts for the transformation to radial coordinates. Next, to obtain the scaling form for the displacement correlations in real space at large $r$, we can use the identity
\begin{eqnarray}
 \int_{1}^{\infty} d k \frac{e^{-i k \mathcal{R}}}{k} &=& \left(-\log \mathcal{R} - \gamma + \frac{\mathcal{R}^2}{4} \right) + \mathcal{O}(\mathcal{R}^3).
\label{eq_kintegration}
\end{eqnarray}
Here $\gamma = \displaystyle{\lim_{n\to \infty}}\left(-\log n + \sum_{k=1}^n\frac{1}{k}\right)\simeq 0.5772$ is the Euler-Mascheroni constant. The scaling form for the displacement correlations can then be obtained by analyzing the behaviour of the integral $\mathcal{I}^{K}(\lambda,r,\theta,\psi)$ in the large distance limit $r \gg 1$ and $\lambda r \ll 1$. To do this, we perform a variable transformation $\kappa = \frac{k}{\lambda}$ and $\rho = \frac{r}{L}$. The integrals can be expressed in terms of the scaled variables as
\begin{eqnarray}
\mathcal{I}^{K}(\lambda,\rho,\theta,\psi) &=&  \int_{1}^{\frac{\pi}{\lambda}} \frac{\exp(-i2\pi \Lambda \kappa\rho \cos(\theta - \psi))}{\kappa} d\kappa.
\end{eqnarray}
We can extend the limit of the integrals in the $\lambda \to 0$ limit  $\frac{\pi}{\lambda} \to \infty$. We therefore have
\begin{eqnarray}
\mathcal{I}^{K}(\lambda,\rho,\theta,\psi) &=& \int_{1}^{\infty} \frac{\exp(-i2\pi \Lambda \kappa\rho \cos(\theta - \psi))}{\kappa} d\kappa.
\label{eq_k_final_K}
\end{eqnarray}
Next, using the identity in Eq.~(\ref{eq_kintegration}) in the limit of small $\mathcal{R}$, along with $ \mathcal{R} = 2\pi\rho\Lambda \cos(\theta - \psi)$, Eq.~(\ref{eq_k_final_K}) can be expressed as
\begin{small}
\begin{eqnarray}
\nonumber
\mathcal{I}^{K}(\Lambda,\rho,\theta,\psi) &=& \Big[ \frac{1}{4} (2\pi \rho \Lambda)^2 \cos^2(\theta - \psi) - \log(\cos(\theta - \psi)) + \log(2\pi\lambda \rho) - \gamma  \Big].\\
\label{eq_continuumform_K}
\end{eqnarray}
\end{small}
Substituting this form back into Eq.~(\ref{eq_correlation_K_intermediate}) leads to the scaling form for the displacement correlations. We find that the uncorrelated disorder in inter-particle bond stiffness leads to the following long-ranged correlations of the displacement fields
\begin{eqnarray}
\mathcal{C}^{K}_{\mu \nu}(r,\theta) \equiv \eta_K^{2} c^{K}_{\mu \nu}(r/L,\theta),
\label{eq_corr_K}
\end{eqnarray}
with the scaling form given by
\begin{eqnarray}
c^{K}_{\mu\nu}(\rho,\theta) &\approx& \text{const}^{K}_{\mu\nu} - \text{a}^{K}_{\mu\nu}(\theta)\log \rho + \text{b}^{K}_{\mu\nu}(\theta) \rho^{2}.
\label{eq_scaling_collapse_K}
\end{eqnarray}
The values of the coefficients $\text{const}^{K}_{\mu \nu}, \text{a}^{K}_{\mu \nu}(\theta)$ and $\text{b}^K_{\mu \nu}(\theta)$ depend on the indices $\mu \nu$. The coefficients have the following form 
\begin{small}
\begin{eqnarray}
\nonumber
\text{const}^{K}_{\mu \nu} &=& \frac{1}{2 \Lambda^2(2\pi)^4}\int_{-\pi}^{\pi} \left[\log(\cos(\theta-\psi)) - \log(2\pi\Lambda) + \gamma \right ] \\
\nonumber
&&\hspace{2.2cm}\times \sum_{\alpha,\beta} \Big[\Tilde{g}^{\mu \alpha}(\psi)\Tilde{g}^{\nu \alpha}(\psi) s_{K2}^{\alpha \beta}(\psi)\Big]d \psi, \\
\text{a}^{K}_{\mu\nu}(\theta) &=& \frac{1}{2 (2\pi)^2} \int_{-\pi}^{\pi} \sum_{\alpha,\beta} \Big[\Tilde{g}^{\mu \alpha}(\psi)\Tilde{g}^{\nu \alpha}(\psi) s_{K2}^{\alpha \beta}(\psi)\Big]  d\psi, \\
\nonumber
\text{b}^{K}_{\mu\nu}(\theta) &=& \frac{1}{2(2\pi)^2} \int_{-\pi}^{\pi} \cos^2(\theta - \psi) (\pi \Lambda)^2 \sum_{\alpha,\beta} \Big[\Tilde{g}^{\mu \alpha}(\psi)\Tilde{g}^{\nu \alpha}(\psi) s_{K2}^{\alpha \beta}(\psi) \Big]  d\psi.
\label{eq_asymptotic_corr_poly}
\end{eqnarray}
\end{small}
The scaling forms and the predicted logarithmic behaviour as expressed in Eq.~\eqref{eq_scaling_collapse_K} in the $\rho = r/L \to 0$ limit are plotted in Fig.~\ref{fig_corr_K1}.

\section{Polydispersed Soft Particle Crystals}
\label{sec_correlations_poly}
We next analyze the displacement correlations produced by microscopic disorder arising due to polydispersity in particle sizes.
In order to obtain these correlation functions, we utilize the expressions for the displacement fields derived in Eq.~(\ref{eq_disp_fourier}). We introduce disorder in the system by varying the sizes of the particles (see Fig.~\ref{fig_main_schematic}) as 
\begin{eqnarray}
\sigma_i = \sigma_0 (1+ \eta_{p} \xi_i) = \sigma_0 + \delta \sigma_i, 
\label{eq_disorder}
\end{eqnarray}
where $\xi_i$ are independent random variables at each site drawn from a uniform distribution between $[-1/2,1/2]$ with $\sigma_0 = 1/2$ \cite{tong2015crystals,acharya2020athermal}. All the interparticle bond stiffnesses are fixed at $K_{ij} = K_0 = 1$. Here $\eta_p$ represents the polydispersity strength, and the correlations in the incremental size of the particles are therefore $\langle \delta \sigma_i \delta \sigma_j \rangle = (\eta_p^2/12) \sigma_0^2 \delta_{ij}$ \cite{acharya2020athermal}, which in Fourier space yields 
\begin{equation}
\langle \delta{\Tilde{\sigma}}(\vec{k})  \delta{\Tilde{\sigma}}(\vec{k}') \rangle  = (\eta_{p}^2 V/48) \delta(\vec{k} + \vec{k}').  
\label{eq_corr_sigma}
\end{equation}
The equations of force balance along the $x$ and $y$ directions are
\begin{eqnarray}
\nonumber
\sum_j \left(C_{ij}^{xx}\delta x^{(1)}_{ij} + C_{ij}^{xy}\delta y^{(1)}_{ij} + C_{ij}^{x\sigma}\delta \sigma_{ij}\right) &=& 0, ~\forall ~i, \\
\sum_j \left(C_{ij}^{yx}\delta x^{(1)}_{ij} + C_{ij}^{yy}\delta y^{(1)}_{ij} + C_{ij}^{y\sigma}\delta \sigma_{ij}\right) &=& 0, ~\forall ~i.
\label{1st_order_force}
\end{eqnarray}
where $C_{ij}^{x\sigma} = \frac{(R_0 - \sigma_0)}{4\sigma_0^3} \cos\big(\frac{\pi j}{3}\big)$ and $C_{ij}^{x\sigma} = \frac{(R_0 - \sigma_0)}{4\sigma_0^3} \cos\big(\frac{\pi j}{3}\big)$. The source terms at each site associated with the disorder in particle sizes can then be expressed in Fourier space as as~\cite{acharya2021disorder}
\begin{eqnarray}
\nonumber
\tilde{S}^{x}(\vec{k}) &=& -D^{x}(\vec{k}) \delta \tilde{ \sigma}(\vec{k}),\\
\tilde{S}^{y}(\vec{k}) &=& -D^{y}(\vec{k}) \delta \tilde{ \sigma}(\vec{k}),
\label{eq_source_poly_k}
\end{eqnarray}
where $\delta \tilde{\sigma}(\vec{k})$ is fourier transform of $\delta \sigma_i$ and 
\begin{equation}
D^\mu(\vec{k}) = -\sum_{j=0}^{5} \left(1+\exp(i \vec{k}.\vec{\mathbb{r}}_j))\right) C^{\mu \sigma}_{ij}.
\end{equation}
\begin{figure}[t!]
\centering
\includegraphics[width=0.9\linewidth]{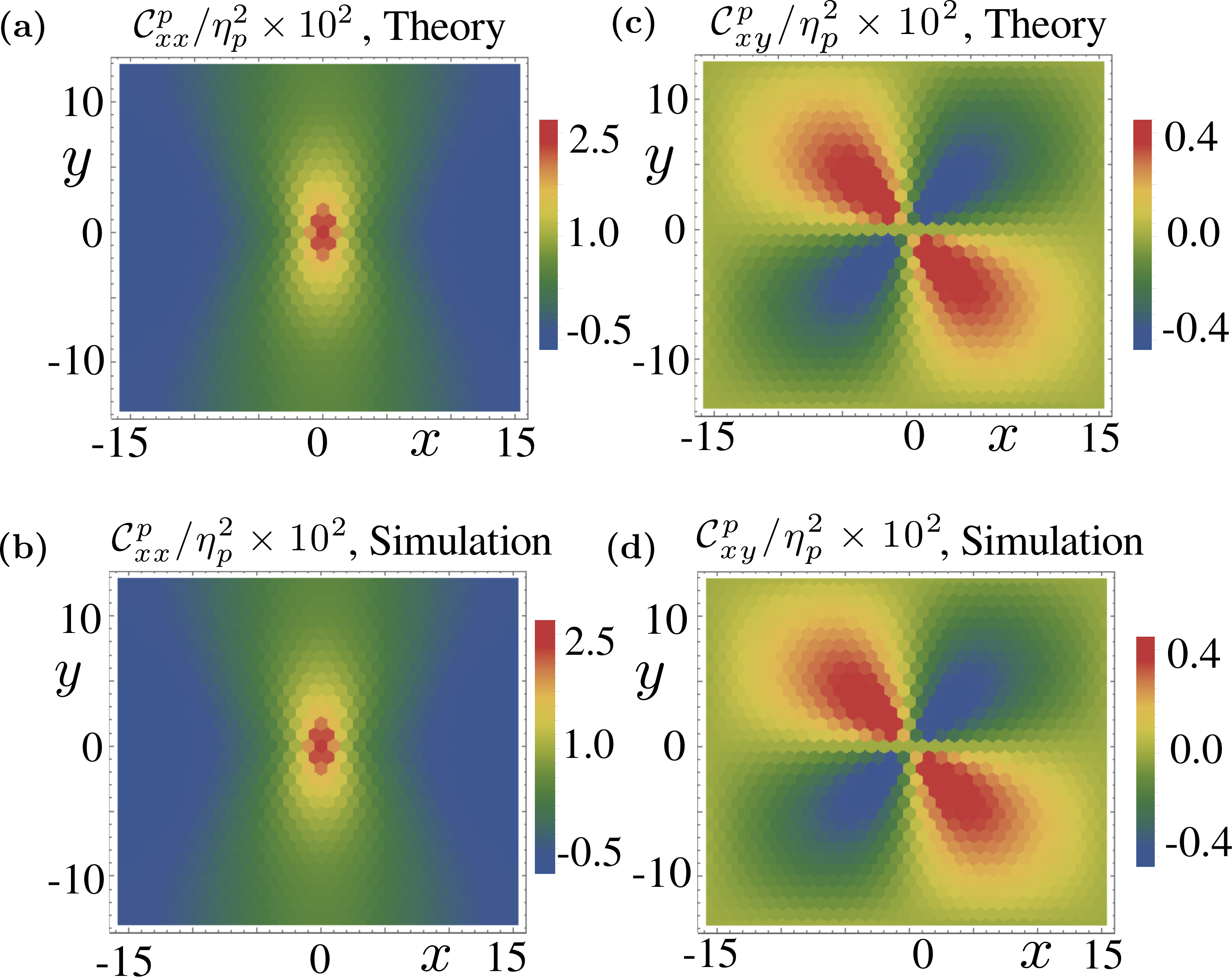}
\caption{Correlations in the displacement fields produced by polydispersity in particle sizes \textbf{(a)} $\mathcal{C}^{p}_{xx}(\vec{r}-\vec{r}' ) = \langle \delta x(\vec{r}) \delta x( \vec{r}' ) \rangle$ obtained from the expression in Eq.~\eqref{eq_correlation_real_poly1}. 
\textbf{(b)} $\mathcal{C}^{p}_{xx}(\vec{r}-\vec{r}' ) = \langle \delta x(\vec{r}) \delta x( \vec{r}' ) \rangle$ obtained from the simulations.
\textbf{(c)} $\mathcal{C}^{p}_{xy}(\vec{r}-\vec{r}' ) = \langle \delta x(\vec{r}) \delta y( \vec{r}' ) \rangle$ obtained from the expression in Eq.~\eqref{eq_correlation_real_poly1}.
\textbf{(d)} $\mathcal{C}^{p}_{xy}(\vec{r}-\vec{r}' ) = \langle \delta x(\vec{r}) \delta y( \vec{r}' ) \rangle$ obtained from simulations.  The magnitude of the disorder used is $\eta_p = 0.001$.
}
\label{Fig_correlation_polydispersity}
\end{figure}
Using Eq.~\eqref{eq_corr_sigma} and Eq.~\eqref{eq_source_poly_k} in the expression in Eq.~\eqref{eq_correlations_exp_K}, we arrive at the correlations in the displacement fields in Fourier space
\begin{small}
\begin{eqnarray}
\hspace{-1cm}
\tilde{\mathcal{C}}^{p}_{\mu \nu}(\vec{k}) &=& \frac{\eta_{p}^2 V}{48} \left[\Big(\sum_{\alpha}\Tilde{G}^{\mu \alpha}(\vec{k}) \tilde{D}^{\alpha}(\vec{k})\Big)\Big(\sum_{\alpha}\Tilde{G}^{\nu\alpha}(-\vec{k}) \tilde{D}^{\alpha}(-\vec{k})\Big)\right].
\label{eq_correlation_poly_k}
\end{eqnarray}
\end{small}
The displacement correlations in real space can then be obtained by taking an inverse Fourier transform. 
Using Eq.~\eqref{eq_correlation1}, we obtain the real space correlations as
\begin{small}
\begin{eqnarray}
\nonumber
\mathcal{C}^{p}_{\mu\nu}(\vec{r}) &=&  \frac{\eta_{p}^2 V}{48} \sum_{\vec{k}} \left[\Big(\sum_{\alpha}\Tilde{G}^{\mu \alpha}(\vec{k}) \tilde{D}^{\alpha}(\vec{k})\Big)\Big(\sum_{\alpha}\Tilde{G}^{\nu\alpha}(-\vec{k}) \tilde{D}^{\alpha}(-\vec{k})\Big)\right]\\
&&\hspace{3cm} \times \exp[i \vec{k}\cdot (\vec{r}-\vec{r}^\prime)].
\label{eq_correlation_real_poly1}
\end{eqnarray}
\end{small}
In Fig.~\ref{Fig_correlation_polydispersity} ({\textbf a}) and Fig.~\ref{Fig_correlation_polydispersity} ({\textbf c}) we plot the correlations $\mathcal{C}_{xx}^p(\vec{r})$ and  $\mathcal{C}_{xy}^p(\vec{r})$ obtained from the above theory, which match with the correlations obtained from simulations (averaging over 1000 {\it random} configurations) plotted in Fig.~\ref{Fig_correlation_polydispersity} ({\textbf b}) and Fig.~\ref{Fig_correlation_polydispersity} ({\textbf d}). The strength of the disorder is set to $\eta_p = 0.001$ in these simulations. 
Comparing Figs.~\ref{fig_corr_K} and ~\ref{Fig_correlation_polydispersity}, it is clear that the correlations produced by random bond stiffness and polydispersity differ in their symmetry properties.

\subsection*{Continuum Limit}
As in Section \ref{sec_bond_disorder}, transforming the sum to an integral in the $L \to \infty$ limit in the inverse Fourier transform of Eq.~(\ref{eq_correlation_poly_k}), the displacement correlation functions can be expressed in real space as
\begin{small}
\begin{eqnarray}
\nonumber
\mathcal{C}^{p}_{\mu \nu }(\vec{r}) &=& \frac{\eta_{p}^{2}}{48(2 \pi)^2} \int_{-\pi}^{\pi} \int_{-\pi}^{\pi}\sum_{\alpha,\beta}\big[(\Tilde{G}^{\mu \alpha}(\vec{k})\tilde{D}^{\alpha}(\vec{k}) \Tilde{G}^{\nu \beta}(-\vec{k}) \tilde{D}^{\beta}(-\vec{k})) \big]\\
&&\hspace{3cm}\times
\exp( -{i \vec{k}.\vec{r}}) d^{2}{\vec{k}}. 
\end{eqnarray}
\end{small}
In the $|k| \to 0$ limit the Green's functions have a simpler form and can be written as $\Tilde{G}^{\mu \nu}(\vec{k}) \sim \frac{g^{\mu\nu}(\psi)}{k^2}$~\cite{acharya2021disorder,das2021long}. Similarly $\Tilde{D}^{\mu}(\vec{k}) $ can be expressed as $\Tilde{D}^{\mu}(\vec{k})\sim\mathcal{D}^{\mu}(\psi) k$~\cite{acharya2021disorder}. Therefore, the total correlation function in Fourier space has the behaviour $\tilde{\mathcal{C}}^{p}_{\mu \nu}(\vec{k}) \sim \frac{\eta_{p}^2}{k^2}$. The displacement correlations can therefore be expressed as
\begin{small}
\begin{eqnarray}
\nonumber
\mathcal{C}^{p}_{\mu \nu }(\vec{r}) &=& \frac{\eta_{p}^{2}}{48(2 \pi)^2} \int_{-\pi}^{\pi} \sum_{\alpha,\beta}\big[(\Tilde{g}^{\mu \alpha}(\psi)\Tilde{g}^{\nu \beta}(\psi)\Tilde{\mathcal{D}}^{\alpha}(\psi) \Tilde{\mathcal{D}}^{\beta}(\psi)) \big] d\psi\\
&&\hspace{3cm}\times
\underbrace{{\int_{\lambda}^{\pi}\frac{\exp( -{i \vec{k}.\vec{r}})}{k} d{k}}}_{\mathcal{I}^{p}(\lambda,r,\theta,\psi)}.
\label{eq_correlation_poly_intermediate}
\end{eqnarray}
\end{small}

Here, $\lambda$ is a cutoff that represents the system-size dependent lower limit in radial fourier coordinates.
Here $\lambda$ represents the system-size dependent lower limit in radial coordinates in Fourier space $\lambda = \frac{2\pi}{L}\Lambda$, and $\Lambda$ represents an $\mathcal{O}(1)$ tuning parameter as explained in Section~\ref{sec_bond_disorder}. The continuum form of the displacement correlations can be obtained by evaluating $\mathcal{I}^{p}(\lambda,r,\theta,\psi)$. Using Eq.~(\ref{eq_kintegration}) in Eq.~(\ref{eq_correlation_poly_intermediate}), the scaling form for the displacement correlations can be calculated using the method described in Section~\ref{sec_bond_disorder}. The uncorrelated disorder in particle sizes then leads to long-ranged correlations in real space
\begin{eqnarray}
\mathcal{C}^{p}_{\mu \nu}(r,\theta) \equiv \eta_p^{2} c^{p}_{\mu \nu}(r/L,\theta),
\label{eq_corr_poly}
\end{eqnarray}
with the scaling form given by
\begin{eqnarray}
c^{p}_{\mu\nu}(\rho,\theta) &\approx& \text{const}^{p}_{\mu\nu} - \text{a}^{p}_{\mu\nu}(\theta)\log \rho + \text{b}^{p}_{\mu\nu}(\theta) \rho^{2}.
\label{eq_scaling_collapse_poly}
\end{eqnarray}

\begin{figure}[t!]
\centering
\includegraphics[width=0.9\linewidth]{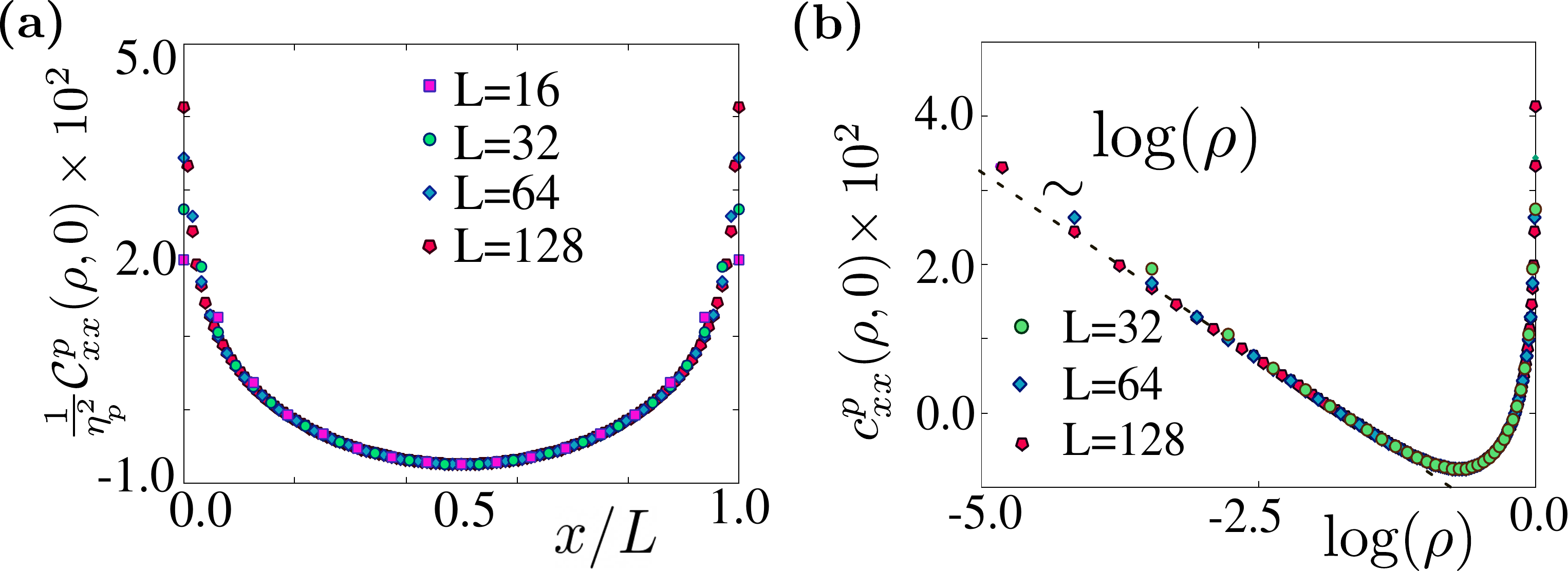}
\caption{\textbf{(a)} $\delta x$ correlations produced by particle size polydispersity, measured along the $x$ direction $\mathcal{C}^{p}_{xx}(x,0) = \langle \delta x(\vec{r}) \delta x(\vec{r}+x\hat{x}) \rangle$, for different system sizes. These correlations follow the scaling prediction in Eq.~(\ref{eq_corr_poly}), with a scaling variable $\rho = r/L$. \textbf{(b)} These correlations display logarithmic behaviour as $\rho \to 0$ consistent with our predictions $\mathcal{C}^{p}_{\mu \nu}(\vec{r}) \sim \log(\rho)$ in Eq.~(\ref{eq_scaling_collapse_poly}).
}
\label{Fig_correlation_polydispersity1}
\end{figure}

We note that the scaling form for the displacement correlations arising due to polydispersity is very similar to that obtained due to disorder in bond stiffness. The values of the coefficients $\text{const}^{p}_{\mu \nu}, \text{a}^{p}_{\mu \nu}(\theta)$ and $\text{b}^{p}_{\mu \nu}(\theta)$ depend on the indices $\mu \nu$. The coefficients have the following form 
\begin{small}
\begin{eqnarray}
\nonumber
\text{const}^{p}_{\mu \nu} &=& \frac{1}{2 \Lambda^2(2\pi)^4}\int_{-\pi}^{\pi} \left[\log(\cos(\theta-\psi)) - \log(2\pi\Lambda) + \gamma \right ] \\
\nonumber
&&\hspace{2.0cm}\times \sum_{\alpha,\beta} \Big[\Tilde{g}^{\mu \alpha}(\psi)\Tilde{g}^{\nu \alpha}(\psi) \tilde{\mathcal{D}}^{\alpha}(\psi) \tilde{\mathcal{D}}^{\beta}(\psi)\Big]d \psi, \\
\nonumber
\text{a}^{p}_{\mu\nu}(\theta) &=& \frac{1}{2 (2\pi)^2} \int_{-\pi}^{\pi} \sum_{\alpha,\beta} \Big[\Tilde{g}^{\mu \alpha}(\psi)\Tilde{g}^{\nu \alpha}(\psi) \tilde{\mathcal{D}}^{\alpha}(\psi) \tilde{\mathcal{D}}^{\beta}(\psi) \Big]  d\psi, \\
\nonumber
\text{b}^{p}_{\mu\nu}(\theta) &=& \frac{1}{2(2\pi)^2} \int_{-\pi}^{\pi} \cos^2(\theta - \psi) (\pi \Lambda)^2 \\
&&\hspace{2.0cm} \times \sum_{\alpha,\beta} \Big[\Tilde{g}^{\mu \alpha}(\psi)\Tilde{g}^{\nu \alpha}(\psi) \tilde{\mathcal{D}}^{\alpha}(\psi) \tilde{\mathcal{D}}^{\beta}(\psi) \Big]  d\psi.
\label{eq_asymptotic_corr_poly} 
\end{eqnarray}
\end{small}
Interestingly, the coefficient multiplying the logarithmic term  $\text{a}^{p}_{\mu\nu}(\theta)$ is independent of the direction $\theta$.  In Fig.~\ref{Fig_correlation_polydispersity1} we plot the theoretically obtained correlations along with the predicted logarithmic behaviour as $\rho = r/L \to 0$.


\section{Triangular Network with Random Pinning Forces}
\label{sec_correlations_random_forces}

We next analyze the correlations in displacement fields of the system in the presence of {\it disordered} external forces. Such situations naturally arise in dense systems with activity where jammed regions arise with the directions and magnitudes of the individual active forces being randomly distributed. \cite{merrigan2020arrested,cates2015motility,rituparno}. 
We assume the relaxation timescale to achieve the mechanical equilibrium is considerably less than the timescale of the angular fluctuations of the pinned forces. This assumption is valid for the case of near-rigid particles. 
In order to simulate such systems, we impose external forces $\delta f^{x(y)}_{\text{ext}}(\vec{r})$ 
at each vertex which are chosen from a delta correlated Gaussian distribution such that 
\begin{eqnarray}
\langle {\delta{f}_{\text{ext}}^{\mu}}(\vec{r})  {\delta{f}_{\text{ext}}^{\nu}}(\vec{r}') \rangle  = \eta_f^2 \delta_{\mu \nu} \delta \left(\vec{r} - \vec{r}' \right).
\label{eq_force_corr}
\end{eqnarray} 
We maintain the global force balance by imposing an additional force at each vertex as discussed in Section~\ref{sec_model}. Therefore in Fourier space, the correlations of the external pinning force can be written as
\begin{small}
\begin{eqnarray}
\label{new_kcorrelation}
\langle {\delta\tilde{
f}_{\text{ext}}^{\mu}}(\vec{k})  {\delta \tilde{f}_{\text{ext}}^{\nu}}(\vec{k}') \rangle  = \eta_f^2 \delta_{\mu \nu} \left( \delta \left(\vec{k}  + \vec{k}' \right) - \frac{\delta(\vec{k}) \delta(\vec{k}')}{L^2}  \right).
\label{eq_Fourier_force_correlations}
\end{eqnarray}
\end{small}
\begin{figure}[t!]
\centering
\includegraphics[width=0.9\linewidth]{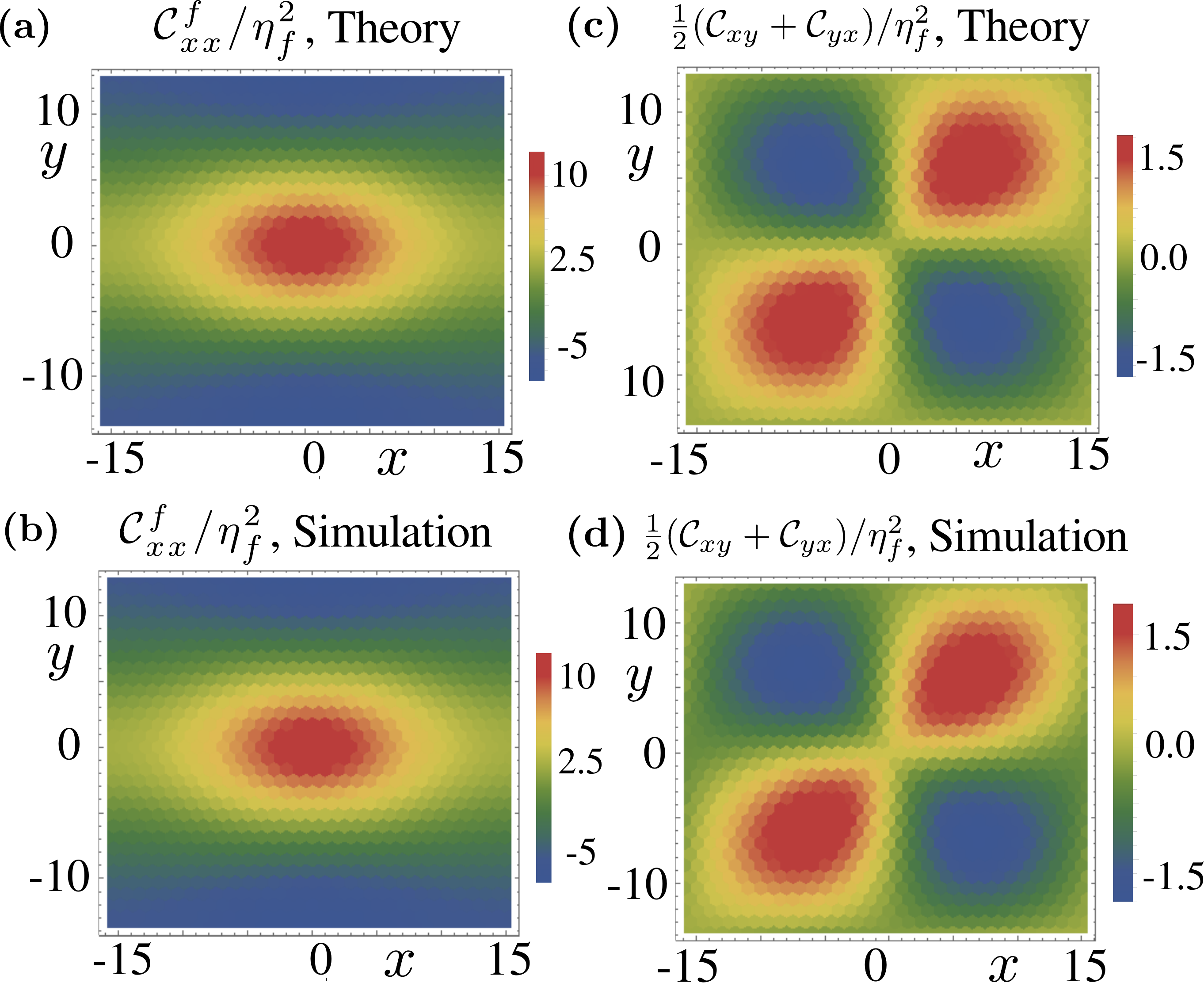}
\caption{Correlations in the displacement fields produced by uncorrelated pinned forces at each site \textbf{(a)} $\mathcal{C}^{p}_{xx}(\vec{r}-\vec{r}' ) = \langle \delta x(\vec{r}) \delta x( \vec{r}' ) \rangle$ obtained from the expression in Eq.~\eqref{eq_correlation_real_force1}. 
\textbf{(b)} $\mathcal{C}^{p}_{xx}(\vec{r}-\vec{r}' ) = \langle \delta x(\vec{r}) \delta x( \vec{r}' ) \rangle$ obtained from simulations.
\textbf{(c)} $\mathcal{C}^{p}_{xy}(\vec{r}-\vec{r}' ) = \langle \delta x(\vec{r}) \delta y( \vec{r}' ) \rangle$ obtained from the expression in Eq.~\eqref{eq_correlation_real_force1}.
\textbf{(d)} $\mathcal{C}^{p}_{xy}(\vec{r}-\vec{r}' ) = \langle \delta x(\vec{r}) \delta y( \vec{r}' ) \rangle$ obtained from simulations. The magnitude of the disorder used is $\eta_f = 0.0001$.
}
\label{Fig_correlations_force}
\end{figure}
Using the above expression in Fourier space the correlation in displacement fields can be written as
\begin{small}
\begin{eqnarray}
\label{eq_force_correlationk1}
\tilde{\mathcal{C}}^{f}_{\mu \nu}(\vec{k}) &=&  \eta_{f}^2 \left[\sum_{\alpha} \Tilde{G}^{\mu \alpha}(\vec{k})\Tilde{G}^{\nu\alpha}(-\vec{k})\right],
\end{eqnarray}
\end{small}
where 
$\alpha \equiv (x,y)$. The real space correlations can then be obtained using Eq.~\eqref{eq_correlation1}. We have
\begin{small}
\begin{eqnarray}
\mathcal{C}^{f}_{\mu\nu}(\vec{r}) &=&  \eta_{f}^2 \left[\sum_{\alpha} \Tilde{G}^{\mu \alpha}(\vec{k})\Tilde{G}^{\nu\alpha}(-\vec{k})\right]\exp[i \vec{k}\cdot (\vec{r}-\vec{r}^\prime)].
\label{eq_correlation_real_force1}
\end{eqnarray}
\end{small}
We plot these correlations in Fig.~\ref{Fig_correlations_force}, along with the correlations obtained from simulations for a disorder strength $\eta_f = 0.0001$, displaying an exact match. Comparing Figs.~\ref{Fig_correlations_force} and \ref{fig_corr_K} with Fig.~\ref{Fig_correlation_polydispersity}
it is clear that the displacement correlations arising due to random pinning forces at each site as well as bond disorder possess different symmetries to those induced by polydispersity in particle sizes. As the displacement correlations are related to the elastic response~\cite{didonna2005nonaffine}, we conclude that the elastic properties arising due to polydispersity are very different from those induced by the presence of random forces at each site.

\subsection*{Continuum Limit}
In the $|k| \to 0$ limit the displacement correlations can be expressed as $\tilde{\mathcal{C}}_{\mu \nu}(\vec{k}) \sim \tilde{g}^{\mu\nu}(\psi)\frac{\eta_{f}^2}{k^4}$.
The correlation functions in real space can be computed by performing an inverse Fourier transform and display long-range behaviour given by
\begin{eqnarray}
\mathcal{C}^{f}_{\mu \nu}(r,\theta) \equiv V \eta_f^{2} c^{f}_{\mu \nu}(r/L,\theta),
\end{eqnarray}
where
\begin{eqnarray}
c^{f}_{\mu \nu}(\rho, \theta) &\approx& \text{const}^{f}_{\mu \nu} -( \text{a}^{f}_{\mu \nu}(\theta) + \text{b}^{f}_{\mu \nu}(\theta) \log \rho ) \rho^2.
\label{eq_scaling_collapse}
\end{eqnarray}
Here the coefficients $\text{const}^{f}_{\mu \nu}, \text{a}^{f}_{\mu \nu}(\theta)$ and $\text{b}^{f}_{\mu \nu}(\theta)$ depend on the indices $\mu \nu$ \cite{das2021long}.
The coefficients can be expressed as
\begin{eqnarray}
\nonumber
\text{const}^{f}_{\mu \nu} & = & \frac{1}{2 \Lambda^2(2\pi)^4}\int_{-\pi}^{\pi} \sum_{\alpha} \Big[\Tilde{g}^{\mu \alpha}(\psi)\Tilde{g}^{\nu \alpha}(\psi) \Big], \\
\nonumber
\text{a}^{f}_{\mu\nu}(\theta) & = & \frac{1}{2(2\pi)^2} \int_{-\pi}^{\pi} \left(\log(\mid \cos(\theta-\psi) \mid)+ \log(2\pi\Lambda) + \gamma - \frac{3}{2}\right) \\
\nonumber
&&\times \cos^2(\theta - \psi) \sum_{\alpha} \Big[\Tilde{g}^{\mu \alpha}(\psi)\Tilde{g}^{\nu \alpha}(\psi) \Big]  d\psi,\\
\text{b}^{f}_{\mu\nu}(\theta) & = & \frac{1}{2 (2\pi)^2}\int_{-\pi}^{\pi} \cos^2(\theta - \psi) \sum_{\alpha} \Big[\Tilde{g}^{\mu \alpha}(\psi)\Tilde{g}^{\nu \alpha}(\psi) \Big]  d\psi.
\end{eqnarray}

\begin{figure}[t!]
\centering
\includegraphics[width=0.9\linewidth]{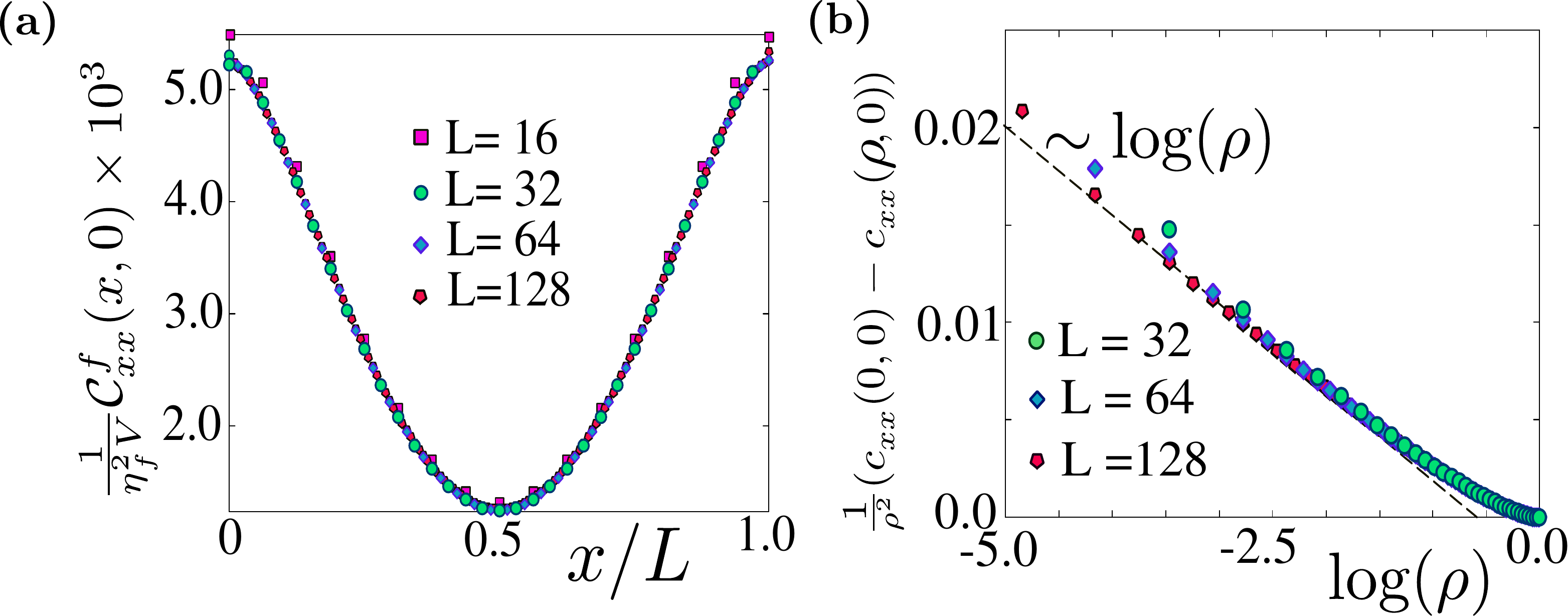}
\caption{\textbf{(a)} $\delta x$ correlations produced by random pinning forces at every site, measured along the $x$ direction $\mathcal{C}^{f}_{xx}(x,0) = \langle \delta x(\vec{r}) \delta x(\vec{r}+x\hat{x} ) \rangle$, for different system sizes. These correlations follow the scaling prediction in Eq.~(\ref{eq_scaling_collapse}), with a scaling variable $\rho = r/L$. \textbf{(b)} These correlations display a logarithmic correction as $\rho \to 0$, consistent with our predictions $\mathcal{C}^{f}_{\mu \nu}(\vec{r}) \sim \rho^2 \log(\rho)$ in Eq.~(\ref{eq_scaling_collapse}).
}
\label{Fig_correlations_force1}
\end{figure}

These displacement correlations exhibit a non-trivial system size scaling. In addition, the correlations also diverge as the volume of the system. Additionally, the correlation functions arising due to random external forces also differ in their asymptotic behaviour from the correlations produced due to polydispersity and random bond stiffness.
The above displacement correlations exhibit a non-analytic behaviour $\sim \rho^2 \log(\rho,\theta)$ as $\rho = r/L \to 0$,
which is different from the nature of the displacement correlations obtained from polydispersity and random bond stiffness, where the additional $\rho^2$ term is absent. This difference can be attributed to the fact that in the case of disorder in bond stiffness and particle sizes, the corresponding source terms at each site have contributions from the disorder at their neighbouring sites.      
We plot these correlations along with the scaling forms and the predicted logarithmic correction as $\rho = r/L \to 0$ in Fig.~\ref{Fig_correlations_force1}. 

\section{Randomly Pinned Polydispersed Crystals}
\label{sec_correlations_both}
Finally, we turn our attention to the displacement correlations in randomly pinned disordered networks. Examples of such pinned forces arise frequently in biological contexts where pulling forces are generated in contractile polymer networks, and pushing forces are generated by the polymerization of polymer networks~\cite{schwarz2013physics}. Such forces are intricately linked to the material properties of the cellular constituents~\cite{pinto2012actin,kruse2000actively}. In this context, it becomes important to study the effects of external pinning forces on disordered backgrounds. 

\begin{figure}[t]
\centering
\includegraphics[width=0.9\linewidth]{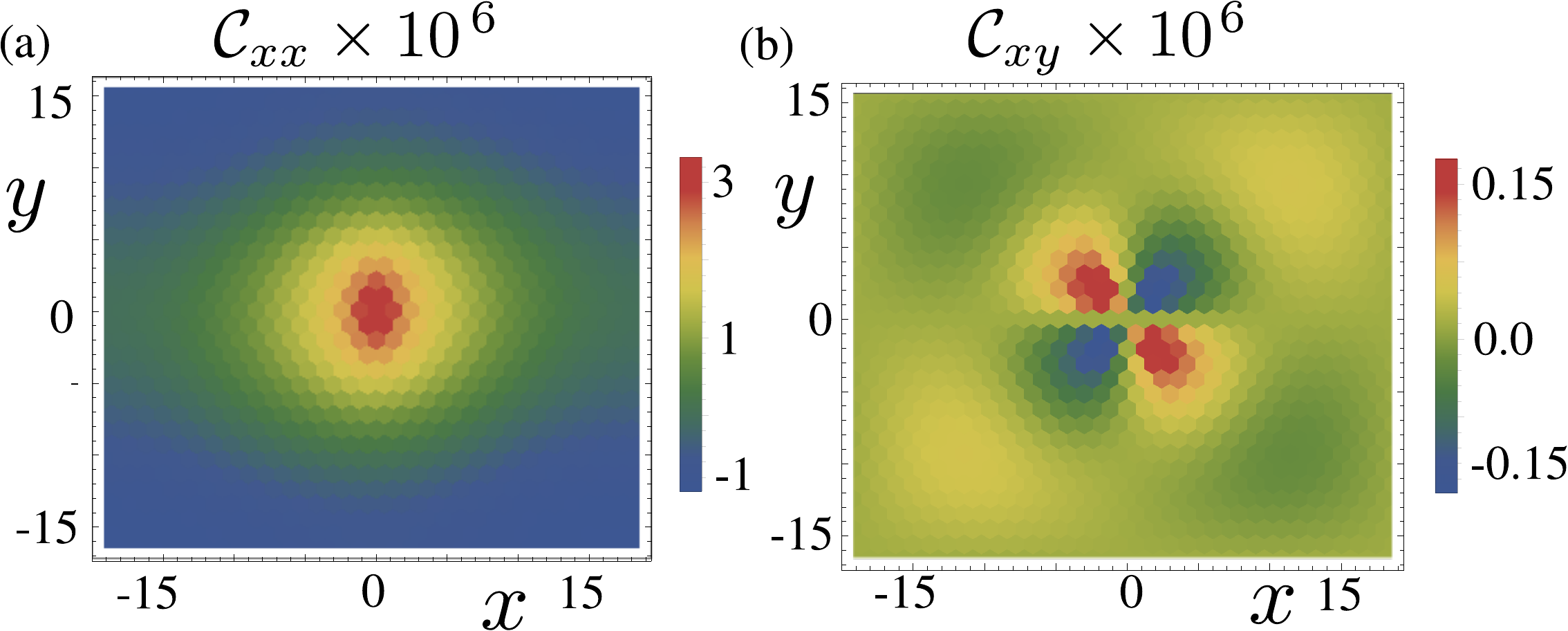}
\caption{Correlations in the displacement fields produced by random pinning forces (with $\eta_f = 0.0004$) in a network with polydispersity in particle sizes (with $\eta_p = 0.0076$), obtained using Eq.~(\ref{eq_corr_both}). These correlations are a superposition of the correlations arising from each disorder individually \textbf{(a)} $\mathcal{C}_{xx}(\vec{r}-\vec{r}' ) = \mathcal{C}^{p}_{xx}(\vec{r}-\vec{r}') + \mathcal{C}^{f}_{xx}(\vec{r}-\vec{r}' )$, \textbf{(b)} $\mathcal{C}_{xx}(\vec{r}-\vec{r}' ) = \mathcal{C}^{p}_{xx}(\vec{r}-\vec{r}') + \mathcal{C}^{f}_{xx}(\vec{r}-\vec{r}')$.
}
\label{Fig_correlation_both}
\end{figure}

Since we are interested in the response of the system at linear order, the displacement fields produced by the presence of multiple types of disorder can be expressed as a linear combination of the displacement fields arising from the individual source terms corresponding to each disorder individually. As the individual source terms are uncorrelated, the displacement correlations are a superposition of the correlations arising from each individual disorder. The total displacement correlation as a response to random external forces on a disordered background arising due to polydispersity in particle sizes can be computed as
\begin{small}
\begin{eqnarray}
\Tilde{\mathcal{C}}_{\mu \nu}(\vec{k}) &=& \Tilde{\mathcal{C}}^{f}_{\mu \nu}(\vec{k}) + \Tilde{\mathcal{C}}^{p}_{\mu \nu}(\vec{k}),
\label{eq_corr_both}
\end{eqnarray}
\end{small}
where the individual correlations are given in Eqs.~\eqref{eq_correlation_poly_k} and ~\eqref{eq_force_correlationk1} respectively.
This leads to the non-trivial prediction of the displacement correlations depicted in Fig.~\ref{Fig_correlation_both}, where we plot the displacement correlations obtained theoretically using the expressions in Eq.~\eqref{eq_corr_both}. Numerically obtained correlations using energy minimized configurations of polydispersed disks with external pinning forces match the above theory exactly.

As explained in Sections~\ref{sec_correlations_poly} and ~\ref{sec_correlations_random_forces}, the displacement correlations arising from the two types of disorder individually display long-ranged behaviour. However, they differ in their symmetry properties as is clear from Figs.~\ref{Fig_correlation_polydispersity} and ~\ref{Fig_correlations_force}. Therefore, when both types of disorder are present, the displacement correlations exhibit exotic forms in regions of the parameter space $\{ \eta_f,\eta_p \}$ as shown in Fig. \ref{Fig_correlation_both}. This can lead to non-trivial lengthscales at which crossovers between different symmetries in the displacement correlations occur in such systems.
In addition, since the correlations arising from external forces diverge with the volume of the system, we expect that for large enough system sizes, the displacement correlations are dominated by the pinned random forces.

\section{Conclusions and Discussion}
\label{sec_conclusions}
In this paper we have developed a formalism to exactly compute displacement correlations in disordered athermal networks in the presence of small microscopic disorder. 
We introduced a generalized disordered network model which we have used to compute displacement correlations for various physically relevant limits, such as introducing polydispersity in particle sizes or incorporating randomness in the bond stiffness. Our formalism allows us to utilize the disorder at the microscopic scale to derive the exact displacement fields at each vertex. This enables a derivation of the exact correlation functions on a disordered background.
Our analytic results demonstrate that {\it uncorrelated} disorder at the microscopic scale generates long-range correlations in athermal systems, as has also been observed within a field theoretic framework \cite{ro2021disorder}. 
Our analysis also demonstrates that long-range correlations arising from disorder in bond stiffness differ in symmetry properties from correlations produced by polydispersity. 
We also introduced external disorder by incorporating random forces at each vertex of the network. We demonstrated that displacement correlations produced by external disorder displays different scaling properties in comparison to the correlations produced by internal disorder such as polydispersity and random bond stiffness.
Additionally, the non-analytic logarithmic corrections arising in the scaling limit of the correlations do not arise within a continuum elasticity framework.
We also showed that the correlations produced by multiple types of disorder can be computed as a superposition of the individual correlations, and demonstrated this for a polydispersed athermal membrane with random external forces imposed at each vertex.
In addition, we demonstrated that the response of networks to disorder induced by polydispersity in particle sizes is very different from the response arising from uncorrelated pinning forces, allowing for exotic forms of the transverse correlation functions, when both types of disorder are present. 


The results presented in this paper can be used as a starting point to build theories for networks where geometric disorder plays an important role. For example, one can analyze the change in the vibrational density of states induced by the different microscopic disorders in the system. Our theory can also be useful in understanding the stability of near-crystalline systems to different external conditions, such as uniform strain and shear. It would be interesting to extend the formulation presented in this paper to incorporate the effects of friction and non-holonomic constraints such as Coulomb inequalities. Another avenue that can be explored are the correlations in the components of the stress tensor in such near-crystalline disordered materials in order to understand the emergent elasticity and their associated gauge theories \cite{jishnu}. Finally, it would also be interesting to develop these predictions to second order in our perturbation expansion, where the non-linear coupling \cite{acharya2021disorder} between the different disorders can be probed.






\section{Acknowledgments}

We thank \mbox{Surajit Sengupta}, Pinaki Chaudhuri, Mustansir Barma and Bulbul Chakraborty for useful discussions. This project was funded by intramural funds at TIFR Hyderabad from the Department of Atomic Energy (DAE).










\printbibliography
\end{document}